\documentclass[aps,twocolumn,preprintnumbers,amsmath,amssymb,floatfix,groupedaddress,nofootinbib]{revtex4}

\usepackage{enumerate}
\usepackage{graphicx}
\usepackage{color}
\usepackage[utf8]{inputenc}
\usepackage{enumitem}
\usepackage{appendix}

\usepackage{fontenc}  
\usepackage{amsmath}

\usepackage[francais]{babel}

\newcommand{\beq}{\begin{equation}}
\newcommand{\eeq}{\end{equation}}
\newcommand{\bea}{\begin{eqnarray}}
\newcommand{\eea}{\end{eqnarray}}
\newcommand{\ba}{\begin{array}}
\newcommand{\ea}{\end{array}}

\newcommand{\bef}{\begin{figure}}
\newcommand{\eef}{\end{figure}}

\begin{document}

\title{Revisiting Quantum Contextuality. }

\author{Philippe Grangier}

\affiliation{ \vskip 2mm
Laboratoire Charles Fabry, Institut d'Optique Graduate School,  
Centre National de la Recherche Scientifique, Universit\'e Paris~Saclay, F91127 Palaiseau, France. 
philippe.grangier@institutoptique.fr
}

\begin{abstract}
The purpose of this note is to complete the interesting review on quantum contextuality \cite{qc}  that appeared recently. In particular we will introduce and discuss the ideas of extracontextuality and extravalence, that allow one to relate Kochen-Specker's and Gleason's theorems, and also to shift the emphasis from the first to the second one. We will also argue that whereas Kochen-Specker's is essentially a negative result (a no-go theorem), Gleason's is a positive one since it provides a mathematical justification of Born's rule. The link between these issues is provided by a specific quantum feature that we call extravalence. 

\end{abstract}

\maketitle

\section{What is quantum (non)contextuality ?}

For the clarity of the presentation, we will introduce quantum contextuality in a simple way, based upon usual textbook quantum mechanics \cite{proj}. Then we will open some questions, and propose a way to answer them. Finally, we will draw some conclusions from our approach. 

Within basic textbook quantum mechanics (QM), let us consider a complete quantum measurement of a quantity $A$, where ``complete'' means with no degeneracy left;  usually such a measurement is associated with a complete set of commuting operators (CSCO), that we summarize in a single ``observable'' operator  $\hat A$.   Denoting as $a_i$ and $| u_i \rangle$ the eigenvalues and eigenstates of $\hat A$ , one has from the spectral theorem  $\hat A = \sum_i \; a_i \; | u_i \rangle \langle u_i |$  with standard Dirac's notations.  
In the following $\hat A$ will be identified with  a {\bf context}, that can be understood with different meanings : on the physical side the context corresponds to a macroscopic and operational device, able to measure the quantity $A$, whereas on the mathematical side it corresponds to the  observable $\hat A$, or also to the above set of rank-one orthogonal projectors $\{| u_i \rangle \langle u_i | \}$, where $i$ goes from 1 to $N$, the dimension of the Hilbert space. In usual QM these physical and mathematical objects correspond to each other, and we will come back to that in the last part of the paper.  Note that in our simple approach, like in textbook QM,  there is no ``observer", or ``agent", only physical objects made of systems within contexts. 

According again to standard QM, a measurement of $\hat A$ will give one of the eigenvalues $a_i$, whereas other complete measurements $\hat B$, $\hat C$... associated with other contexts, will similarly give one of their eigenvalues $b_j$, $c_k$...  In each context it is clear that the corresponding observable gets one value among $N$ possible ones, so one says that the value $a_i$ is assigned to the observable $\hat A$. 

Now, what is quantum contextuality, in its simplest definition ? It is the observation that in QM, it is impossible to assign simultaneously values  with certainty to all observables in all possible contexts, each one seen as the experimental background of a measurement.  This simple observation clashes with classical physics, where such an assignment is considered essential to describe the laws of nature. The formal proof of quantum contextuality is given by the Kochen-Specker (KS) theorem, showing a contradiction between QM and all models attributing non-contextual values to some (well chosen)  set of observables.  Such a result  is ``negative'', or in  other terms it is a no-go theorem, showing that a large class of non-contextual models (typically the models with non-contextual hidden variables) contradict QM predictions and experimental observations \cite{qc}.  

However, it does not tell much about what is actually happening in QM : the whole purpose of the KS discussion is to tell what QM is not, rather than to tell what QM is -- which is the real issue we are interested in.  

As a step in this direction, we may ask the question : if it is impossible to assign simultaneously values  with certainty to all observables in all possible contexts, how to make sense of what is being observed and measured ? There is an answer from observation : given a result in a context, one cannot in general specify results in other contexts, but one can specify the {\bf probabilities} of such results.  Even more interestingly, the assignment of probabilities to the value of all observables (that is, to  any given measurement result) turns out to be non-contextual ! In other words, all probabilities corresponding to all other results within all  contexts can be simultaneously defined, albeit via a probability rule, and not a certainty rule.

Contrary to the previous negative result, this is major asset : as we will show in more details below, it is the basis of Gleason's theorem, mathematically establishing  Born's rule. It is therefore a ``positive'' result, much more interesting and powerful than Kochen-Specker's theorem. On the mathematical side,  we note also that Kochen-Specker's theorem can be seen as a corollary of Gleason's one, whereas the reverse is not true \cite{qc}. We will come back to these points below. 

Before that, it is important to note that there is a special case for probabilities: if we restrict ourselves only to certainties (probabilities $p$ = 0 or $p$ = 1),  they are actually values (eigenvalues of projection operators), not probabilities (average values of projection operators). Certainties are therefore contextual, as it is not possible to assign truth values (0 or 1) to all measurement results, this is another useful way to see Kochen-Specker's theorem. But on the other hand one can assign non-contextual probabilities as said above, with values given by Born's law or Gleason's theorem. 

As a conclusion of this introduction, QM is both contextual (for the measurement results) and non contextual (for the probabilities assigned to these results). Though this sounds hardly comprehensible, everybody agrees on that, and this is the current status of quantum contextuality as presented in \cite{qc}. 
Our purpose here is to to make sense of this situation, by introducing a few additional ideas and definitions. 

 \vspace{-2mm}
\section{What is quantum (extra)contextuality ? }

In this section we will introduce some new ideas, still within the scope of the usual textbook QM. Then at some point we will diverge, as it will be seen below. 

First, let us introduce a distinction between a usual pure quantum state $| \psi \rangle$, described as  a vector in the relevant  Hilbert space $\cal E$ for the considered system, and the same vector  considered as an eigenstate of a complete measurement, therefore within a given context as introduced above.  We will define a {\bf modality} as the association of $| \psi \rangle$ and the context $\hat A$. By construction a  modality is certain and repeatable : the same $| \psi \rangle$, and the associated eigenvalue, will be found again and again as long as the same $\hat A$ is measured on the same system.  
\vskip 2mm

From this definition one has   
$| \psi  \rangle = | u_i  \rangle$, where $| u_i  \rangle$ is an eigenstate of $\hat A$, and the modality  ``$| \psi  \rangle \equiv  | u_i  \rangle$ considered as an eigenvector of $\hat A$'' will be denoted as $| \psi  \rangle_A$. It is clear that the same vector $| \psi  \rangle$ may appear as an eigenstate of many other observables (actually, an infinity), as soon as the dimension of $\cal E$ is at least 3.  Since the changes of context are continuous, both physically and mathematically, there is actually an infinity of such observables. So there will be many modalities, e.g. $| \psi  \rangle_A$, $| \psi  \rangle_B$, $| \psi  \rangle_C$, all associated with the same $| \psi  \rangle$ : in the langage of usual QM, they {\it are} the same $| \psi  \rangle$; but here they are different modalities. 
So now we start moving away from the usual  language: let us consider that the different modalities $| \psi  \rangle_A$, $| \psi  \rangle_B$, $| \psi  \rangle_C$  are indeed different, though they are all associated with the same vector $| \psi  \rangle$, or equivalently the same projector $| \psi  \rangle  \langle \psi |$ (projectors are actually more suitable for reasons that will appear below). Clearly this creates an equivalence class between modalities, that we will call an extravalence class, where all modalities are associated with the same $| \psi  \rangle$.  This means that certainty can be transferred between contexts, because $| \psi  \rangle$ is an eigenstate in all the corresponding contexts, and the associated physical phenomenon will be called extracontextuality.  
\vskip 2mm

So far so good, but does this help with the previous annoying statement that QM is both contextual for measurement results,  and non contextual for probabilities assignments ? To see  that consider the following statements 

(i) when making measurements, certainty and reproducibility are warranted for a modality in a given context,  and are also warranted for extravalent modalities when changing context. But they are not warranted beyond that, due to contextuality in the assignment of measurement results as established by the KS theorem. 

(ii) in other cases, when changing the context from $\hat A$ to $\hat  B$, the result is probabilistic and  given by Born's rule 
$| \langle \phi | \psi \rangle |^2$ for an initial modality $| \psi  \rangle_A$ and final one $| \phi  \rangle_B$. This probability depends only on the extravalence classes of the initial and final modalities, in agreement with the non-contextuality in the assignment of probabilities, which is a basic hypothesis of Gleason's theorem. 
\vskip 2mm

So by attributing the vector $| \psi \rangle$ not to a ``quantum state", but to an extravalence class of modalities, one gets a more transparent picture of what QM is telling us : certainty and reproducibility do exist, not only within a context, but also within an extravalence class; when changing context and extravalence class,  probabilities are needed, but they are non-contextual because they connect extravalence classes, whatever the contexts are. 

This argument  tells the ``how", but not yet the ``why" : what would be needed is to explain, or at least justify, the origins of statements (i) and (ii).  This can be done within the framework called CSM, for Contexts, Systems and Modalities \cite{csm1,trsa,csm4b,myst}, and we will briefly summarize here some elements taken from \cite{csm4b}.  In this framework systems within contexts are defined as basic objects,  and modalities are properties associated with certainties, in agreement with the first part of statement~(i). Its second part, i.e. the  contextuality of the assignment of measurement results, appears as a consequence of a quantization postulate : for a given system, the maximum number of mutually exclusive modalities is bounded to $N$; this requires a probabilistic description when contexts and extravalence classes are being changed  \cite{csm4b}.  

To get the probability law (ii), one assumes that the probabilities when changing contexts depend only on the extravalence class, and one attributes a projector $| \psi  \rangle  \langle \psi |$ to each extravalence class. These are strong assumptions, but they do agree with empirical evidence, and are justified in detail  in \cite{csm4b}, based on induction. 
Then it is easy to check that all hypotheses for Gleason's theorem are satisfied, and as a consequence, that Born's rule is justified as the only acceptable probability law  \cite{csm4b}. This argument can be made even more convincing by using Uhlhorn's theorem in addition to Gleason's \cite{csm4c}.

We see now that the consequences drawn from extracontextuality are considerably stronger than the ones drawn from contextuality alone : we not only get no-go theorems, but we reach the correct probabilistic structure of QM, essentially by providing a physical justification of Gleason's hypotheses.  This automatically ensures that  Kochen-Specker's theorem is true also, as a corollary of Gleason's, without the need of examining many different scenarii in many different dimensions  \cite{qc}.  

 \vspace{-2mm}
 \section{More about KS contradictions.}
 
 In the previous sections we identified a context with a CSCO, and the CSCO with a single non-degenerate observable $\hat A$.  However in practice most CSCO are sets of co-commuting observables $(\hat A_1, \hat A_2, \hat A_3....)$, each operator having some degenerate eigenvalues, with degeneracy lifted by considering all of them. In a  textbook CSCO, one stops adding operators when all degeneracies are lifted, e.g. with 3 operators, and then $\hat A \equiv (\hat A_1, \hat A_2, \hat A_3)$. But one may keep adding co-commuting operators, staying thus in the same context. This idea is used in many contextuality theorems, a famous example being the  Peres-Mermin square nicely described in \cite{qc} : for two spins $1/2$, $( \hat \sigma_z \otimes I, I \otimes  \hat \sigma_z)$ is a CSCO, but one rather considers the overcomplete set of co-commuting operators  $( \hat \sigma_z \otimes I, I \otimes  \hat \sigma_z, \hat \sigma_z \otimes  \hat \sigma_z)$. 
 There are other exemples where contexts are given as usual CSCO, for instance the well-known ``colored" set of 9 contexts in dimension 4, introduced in \cite{cabello}, and discussed from the CSM point of view in \cite{trsa}.  Another example is the Mermin version of the GHZ argument \cite{mermin}, also discussed in \cite{inference}.  
 
 Obviously there are infinitely many variants of such KS-type contradictions, that may or may not depend on considering a particular initial state, and may involve a great variety  of measurements and inequalities \cite{qc}. This is a quite interesting zoo to explore, with a lot of combinatorics, but one may wonder what is the ultimate lesson it provides. In our view it may be more fruitful to move straight to Gleason's theorem by considering  continuously varying contexts, in agreement with empirical evidence, and to turn our attention on how to integrate both systems and contexts in a unified mathematical description, maybe along the lines proposed in \cite{completing}. 
 \vskip 2mm
 
Contextuality has also some aspects going beyond QM, not as a claim that everything should be quantum, but as a general feature of some probabilistic theories \cite{AK,ED}; elucidating the quantum-classical boundary in such theories is also an interesting question.

\vspace{-2mm}
\section{Conclusion and future directions.}

\vspace{-1mm}
Our conclusion is that, for the purpose of telling what QM is rather than what QM is not,  the crucial feature of quantum contextuality is extracontextuality. Extracontextuality and subsequently extravalence tell ``how much non-contextual'' QM can be, given that it is neither non-contextual (as it would be the case in classical physics) nor fully contextual (then there would be no connection between different contexts, and thus no theory at all).  
 \vskip 2mm
 
Another conclusion is that the usual state vector $| \psi  \rangle$ is incomplete indeed, not due to any ``hidden variable'', but because it gives access to an actual  physical modality $| \psi  \rangle_A$ {\bf only when the context has been specified}.  Correspondingly, $| \psi  \rangle$ can be turned into a non-trivial ($ p \neq 0,1$) probability distribution only within a given measurement context, not admitting $| \psi  \rangle$ as a modality (otherwise one has again $p$ = 0 or 1). This feature implies that $| \psi  \rangle$ is predictively incomplete  \cite{JJ}, and allows the violation of Bell's inequalities, without requiring any nonlocality at the elementary level \cite{inference}. 
 \vskip 2mm
 
On the other hand, the modalities $| \psi  \rangle_A$, $| \psi  \rangle_B$, $| \psi  \rangle_C$  are indeed complete, with deep reaching consequences on all the usual ``paradoxes" of QM. For instance,  in the framework of the Einstein-Bohr debate, it means that $| \psi  \rangle$ is incomplete indeed, as claimed by Einstein, Podolsky and Rosen, and that it must be completed by ``the very conditions that allow future predictions'' (i.e. the context, either $\hat A$, or $\hat  B$, or $\hat  C$), as claimed by Bohr \cite{debate}. 

The above ideas emphasize that a well defined quantum property, i.e. a modality, belongs to a system within a context. This is a quite objective statement, but clearly the system and the context are not described in the same way.  A system with $N$ mutually exclusive modalities is associated with a Hilbert space $\cal E$ of dimension $N$, whereas the context (as a CSCO)  is associated with operators acting in $\cal E$.  Such operators are constructed  from macroscopic  data, such as orientations or positions of  the apparatus, and the whole construction makes clear that {\bf both} systems and contexts  are needed, and that there is no proper way to make one ``emerge'' from the other. 

This leads to the (in)famous question of the ``Heisenberg cut" between system and context and of the universality of the quantum description.
 Within the CSM framework there are two ways to answer this question~:
 
 -- The first and simpler one is to postulate that quantum objects are made of systems within contexts \cite{CO2002,csm1}. This fits quite well with usual textbook QM, and allows one to recover Born's rule from Gleason's theorem \cite{csm4b}, as explained above. The main benefit here is that the usual ``quantum paradoxes" just vanish : QM is certainly not classical, but it results from contextual quantization, which fits quite well within the usual physical realism \cite{CO2002,csm1,myst} - though not within {\bf classical} physical realism. 
 
 -- A second way is to include  both systems and contexts in the same description as incommensurable objects, along the lines introduced e.g. by von Neumann in \cite{JvN1939}. More details are given in \cite{completing}, but this approach requires to use algebraic tools that are not included in the textbook QM considered until now. More technically,  textbook QM corresponds to type I algebra in the Murray - von Neumann classification, whereas the approach quoted here requires typically  type III algebra \cite{completing}.  These algebraic tools open a way to include both systems and contexts in a unified framework, but they require in some sense to ``manipulate infinities", which is quite possible mathematically though not popular in physics nowadays. Exploring further such directions may be quite useful to get finally a unified picture of what QM tells us.

\vskip 2mm

{\bf Acknowledgements.} 
The author thanks Franck  Lalo\"e, Roger Balian, Nayla Farouki, Alexia Auff\`eves, and  Ehtibar Dzhafarov for many discussions, not meaning that they endorse whatever is written above.


\begin{thebibliography}{36}
\bibitem{proj} Here we only consider 
 projective measurements (PVM), where, as explained in \cite{qc}, all the notions of compatibility, i.e., joint measurability, nondisturbance, outcome repeatability, etc., are equivalent to the commutativity of the observables. This is the original framework of the discussion on quantum contextuality, and though it has been extended in many other interesting directions  \cite{qc}, it is enough for our current purpose.
\end{thebibliography}
\end{document}